\documentclass[%
reprint,
showpacs,preprintnumbers,
amsmath,amssymb,
aps,
prb,
]{revtex4-1}
\usepackage{amsmath}
\usepackage{graphicx}
\usepackage{dcolumn}
\usepackage{bm}
\usepackage{color}
\usepackage{pdfpages}

\makeatletter
\AtBeginDocument{\let\LS@rot\@undefined}
\makeatother

\begin{document}

\title{Robust high-fidelity DFT study of the lithium-graphite phase diagram}

\author{Vikram Pande}
\affiliation{%
Department of Mechanical Engineering, Carnegie Mellon University, Pittsburgh, Pennsylvania 15213
}

\author{Venkatasubramanian Viswanathan}%
 \email{venkvis@cmu.edu}
\affiliation{%
Department of Mechanical Engineering, Carnegie Mellon University, Pittsburgh, Pennsylvania 15213
}
\affiliation{%
Department of Physics, Carnegie Mellon University, Pittsburgh, Pennsylvania 15213
}

\date{\today}

\begin{abstract}

Graphite is the most widely used and among the most widely-studied anode materials for lithium-ion batteries. With increasing demands on lithium batteries to operate at lower temperatures and higher currents, it is crucial to understand lithium intercalation in graphite due to issues associated with lithium plating.  Lithium intercalation into graphite has been extensively studied theoretically using density functional theory (DFT) calculations, complemented by experimental studies through X-ray diffraction, spectroscopy, optical imaging and other techniques. In this work, we present a first principles based model using DFT calculations, employing the BEEF-vdW as the exchange correlation functional, and Ising model to determine the phase transformations and subsequently, the thermodynamic intercalation potential diagram. We explore a configurational phase space of about 1 billion structures by accurately determining the important interactions for the Ising model. The BEEF-vdW exchange correlation functional employed accurately captures a range of interactions including vdW, covalent and ionic interactions. We incorporate phonon contributions at finite temperatures and configurational entropy to get high accuracy in free energy and potentials. We utilize the built-in error estimation capabilities of the BEEF-vdW exchange correlation functional and to develop a methodological framework for determining the uncertainty associated with DFT calculated phase diagrams and intercalation potentials. The framework also determines the confidence of each predicted stable phase. The confidence value of a phase can help us to identify regions of solid solutions and phase transformations accurately. Given the subtle differences in energy between lithium intercalation into graphite and lithium plating ($<$0.1 eV), we believe such an error estimation framework is crucial to know the reliability of DFT predictions.
\end{abstract}

\maketitle

\section{Introduction}

Graphite is the most ubiquitous anode material used in Li-ion batteries owing to its very low potential, high cycle life and low cost.\cite{hassoun2015review,zhang2011review} Lithium ion batteries used in a variety of applications (e.g. electric vehicles) are required to perform at low temperatures\cite{parker2016survey} and also very high current rates\cite{ahmed2017enabling}. Under these conditions, the major factor impacting cycle life is lithium plating. \cite{vetter2005ageing,zinth2014lithium,waldmann2014temperature} Lithium plating occurs when the anode potential is lower than the lithium redox potential.\cite{bugga2010lithium} To understand the lithium plating phenomenon, we first need to accurately understand the lithium-graphite phase diagram and the open circuit voltage profile. This raises an important question of whether propensity for lithium plating is thermodynamic or kinetic. Lithium intercalation in graphite has been the subject of numerous experimental and theoretical investigations.\cite{dahn1991phase,dahn1995mechanisms,reynier2003entropy,funabiki1998impedance,levi1997mechanism1,levi1997mechanism2,holzwarth1978theoretical,thinius2014theoretical,ceder1999phase} 
 The various phases of lithium intercalation into graphite have been characterized using X-ray diffraction (XRD),\cite{guerard1975intercalation,billaud1996revisited,funabiki1999stage} optical methods \cite{guo2016li} and spectroscopic methods.\cite{sole2014situ} However, due to the high degree of disorder, its difficult to precisely pinpoint the structure for a particular phase in the lithium-graphite phase diagram, especially the low lithium content phases.\cite{dahn1991phase,reynier2003entropy} Identification of the lithium-graphite phases will also enable us in studying the lithium diffusivity in these phases and pin-point kinetic limitations for intercalation. Theoretical studies employing density functional theory calculations have been used to complement the experimental understanding and have painted a clearer picture of lithium-graphite phase diagram.\cite{persson2010thermodynamic,wang2014van,ganesh2014binding,filhol2008phase,okamoto2013density,pevsic2016density,hazrati2014li,kganyago2003structural,tasaki2014density,thinius2014theoretical,perassi2016theoretical,smith2017intercalation,wan2015study}. A variety of exchange-correlation functionals have been employed to determine the stable phases and lead to vastly different intercalation potentials and different stable phases. For example, PBE with added vdW interactions predicts $x =$ 0.3, 0.375, 0.5 and 1 as stable phases while optB88-vdW predicts $x =$ 0.15, 0.1667, 0.1875, 0.25, 0.375, 0.5, 0.8333 and 1 as stable phases.\cite{persson2010thermodynamic,hazrati2014li} We also see significant differences in the voltages predicted with different functionals with a disagreement of 0.25 V for the range x$=$0.5 to 1 and an even larger disagreements of 0.3-0.4 V for x$<$ 0.5.\cite{okamoto2013density,persson2010thermodynamic,hazrati2014li} This disagreement  clearly shows the need to determine the uncertainty associated with DFT calculations for predicting the lithium-graphite phase diagram. It is also important to quantify the confidence of predicting a stable phase using DFT as it is going to have direct impact on the intercalation potential. For high accuracy, it is important to include the small contributions from configurational entropy and finite temperature contributions from phonons. Lastly it has also been shown that at a given $x$ in Li$_x$C$_6$, lithium-graphite exists as a mixture of stage 1, stage 2 and stage 3 compounds.\cite{dahn1990suppression} A simple convex hull approach, as used earlier, will be unable to capture such disorder. Thus, there is a need to develop theoretical methods to identify regions of coexistence of phases more accurately and also the specific phases that exist during coexistence.
 
Graphite comprises of graphene sheets stacked on each other, which are bound by weak van der Waals forces. 

As lithium intercalates in the space between theses sheets, there is an increase in the Li-C covalent interactions along with a decrease in the van der Waals interactions between the graphite sheets.\cite{divincenzo1984p,imai2007energetic} Given that the subtle interplay between these two interactions determines the phase diagram, it is crucial to employ an exchange correlation function that is capable of accurately predicting over a wide range of bonding environments.\cite{hazrati2014li}  BEEF-vdW is designed such that it minimizes prediction error for a range of data sets involving molecular formation energies, molecular reaction energies, molecular reaction barriers, non-covalent interactions, solid state properties, and chemisorption on solid surfaces.\cite{wellendorff2012density} In addition, the functional possesses Bayesian error estimation, which is designed to reproduce known energetic errors by mapping the uncertainties on the exchange-correlation parameters. This capability allows an error estimation capability and has been employed successfully to understand uncertainties associated with reaction rates in heterogeneous catalysis,\citep{medford2014assessing} activity relationships in electrocatalysis,\cite{deshpande2016quantifying,krishnamurthy2018maximal} mechanical properties of solid electrolytes\cite{ahmad2016quantification} and magnetic materials\cite{houchins2017quantifying}. BEEF-vdW correctly predicts the adhesion of two graphene sheets to be stable by 0.07 eV.

In this study, we will build a refined picture for lithium-graphite phase diagram employing density functional theory calculations using BEEF-vdW exchange correlation functional. We use an Ising model to explore the structural phase space of lithium and natural graphite based compounds. The coefficients of the Ising model are used to carry out a rigorous search over the enormously large configurational phase space of 10$^9$ configurations. We explore the large phase space by varying the lithium concentration in different unit cells of graphite of multiple sizes. The BEEF ensemble of energies from DFT calculations are used to train an ensemble of Ising models giving us ensembles for each Ising model coefficient.  We show that the in-plane Li-Li interaction is dominantly electrostatic in nature as interaction strength is inversely proportional to distance between lithium atoms.  We use the ensemble of Ising models to predict an ensemble of Gibbs free energies for all phases including the small contributions from phonons and configurational entropy, which give us a distribution of intercalation potentials. Based on our analysis, we identify the following stable phases in the lithium intercalation diagram viz. Li$_{0.0313}$C$_6$,Li$_{0.0375}$C$_6$,Li$_{0.0417}$C$_6$,Li$_{0.0469}$C$_6$, Li$_{0.05}$C$_6$, Li$_{0.0625}$C$_6$, Li$_{0.0833}$C$_6$ which are stage 4 compounds,Li$_{0.1}$C$_6$ which is a stage 3 compounds, Li$_{0.3}$C$_6$, Li$_{0.3333}$C$_6$, Li$_{0.5}$C$_6$ which are stage 2 compounds and Li$_{0.75}$C$_6$, Li$_{0.8333}$C$_6$ and LiC$_6$ which are stage 1 compounds. We also determine the confidence of predicting a stable phase from DFT which gives us insights on whether in given region of $x$ the intercalation is unique phase or a solid solution of multiple phases. We have derived a very accurate lithium graphite phase diagram which will be very useful for models used to determine lithium plating under different conditions. We also believe the proposed framework of estimating uncertainty and confidence value of phases will be very important in computational investigations for phase diagrams of battery materials.

\section{Methodology}

In this work, we will focus on the electrochemical lithium intercalation in natural graphite given by:
\begin{equation}
\mathrm{x(Li^+ + e^-) + C_6 \rightleftharpoons Li_xC_6}
\end{equation}
The phase transformation of lithium-graphite compounds is determined by the Gibbs free energy change associated with this process given by: 
\begin{equation}
\mathrm{\Delta G = G_{Li_xC_6} -  G_{C_6} - xG_{Li^+} - xG_{e^-}}
\end{equation}
where $\mathrm{G_{Li_xC_6}}$ is the free energy of the lithium-graphite phase, $\mathrm{G_{C_6}}$ is the free energy of the graphite phase, $G_{Li^+}$ is the free energy of the Li ion solvated by the electrolyte and $\mathrm{G_{e^-}}$ is the free energy of the electron at the potential of the graphite electrode. The sum of the free energy of the Li ions and the electrons can be related to the free energy of bulk lithium metal as shown through the reaction:
\begin{equation}
\mathrm{Li^+ + e^- \rightleftharpoons Li_{(s)}}
\label{li+li}
\end{equation}
which gives us the relation $\mathrm{G_{Li^+} + G_{e^-_{U=0 V}} = G_{Li_{(s)}}}$. This is termed as the computational lithium electrode and provides a tractable way to determine the sum of the free energies of Li ion and electron for concerted Li ion-electron transfer reactions.\cite{viswanathan2013li} Through this relation, the free energy of an electron is now calculated relative to the potential of Li/Li$^+$ redox couple, $\mathrm{G_{e^-} = G_{e^-_{U=0 V}} - eU_{Li/Li^+}}$. Thus substituting the relation in Eq. \ref{li+li}, we get:
\begin{equation}
\mathrm{\Delta G = G_{Li_xC_6} - G_{C_6}  - xG_{Li_{(s)}} + x(eU_{Li/Li^+})}
\label{eq4}
\end{equation}
Eq. \ref{eq4} gives the intercalation potential of a particular phase of the lithium-graphite phase diagram. To derive the thermodynamic intercalation potential diagram, we need to consider phase transformation from one stable phase to another as lithium insets into graphite. The potential for phase transformation from Li$_{\mathrm{x_0}}$C$_6$ to  Li$_{\mathrm{{x_1}}}$C$_6$ can thus be expressed as:
\begin{equation}
\mathrm{V = - \frac{G_{Li_{x_1}C_6} - G_{Li_{x_0}C_6} - (x_1-x_0)G_{Li_{(s)}}}{x_1-x_0}}
\end{equation}
To calculate the intercalation potentials, we need to calculate the free energies of the stable phases. For a particular phase Li$_x$C$_6$, we have large number of possible structures due to a number of possible sites that lithium atoms can occupy between the graphene sheets with different free energies. Under a thermodynamic formulation, intercalation will proceed through phases with the minimum free energy. The Gibbs free energy comprises of enthalpy, entropy and zero point energy as given by, $\mathrm{\Delta G = \Delta H - T\Delta S + \Delta ZPE}$.

\subsection{Enthalpy Calculation}

The enthalpy of formation, $\mathrm{\Delta H}$ of a system consists of the internal energy change of the system, $\mathrm{\Delta U}$ and the pressure-volume work, $\mathrm{\Delta PV}$. It has been shown that the pressure volume work is negligible compared to the change in internal energy for lithium-graphite structures.\cite{ceder1997application} The internal energy of all lithium-graphite structures, lithium and graphite is calculated using density functional theory (DFT) calculations. The density functional theory calculations are carried out using GPAW, which is a real space implementation of the projector-augmented wave method.\cite{mortensen2005real,enkovaara2010electronic} The DFT energies are then used to find the enthalpy of formation for each of the structures given by $\mathrm{\Delta H = H_{Li_xC_6} - xH_{Li} - H_{C_6}}$. 

Lithium intercalation in graphite is an interplay between Li-C interactions, Li-Li repulsion and C-C vdW forces. Thus, it is important to choose an appropriate exchange correlation functional for the DFT calculations.  Experiments have shown that lithium intercalates in graphite in stages.\cite{dahn1991phase,reynier2003entropy} A stage `n' lithium-graphite phase implies that there are n layers of graphene between two adjacent Li layers. In this work, we will explore phases with stages 1, 2, 3 and 4. Phases with higher stages are computationally very expensive to calculate due to the large number of atoms in the unit cell.

All DFT calculations were done with BEEF-vdW exchange correlation functional and used a Monkhorst-Pack grid for the Brillouin zone sampling.\cite{monkhorst1976special} A convergence of 5 meV/6C was achieved with $10\times10\times10$ k-point grid for a graphite unit cell (4C atoms). For the lithium-graphite phases, the k-points were appropriately scaled down as per the size of the unit cell for each of the phases, so as to maintain the same level of convergence. A grid spacing, h = 0.18\AA~ and a Fermi-Dirac smearing width of 0.05 eV was used.

The graphite AA and AB stacking structures were considered and the AB stacking structure is more stable by 0.09 eV compared to the AA stacking structure, consistent with experiments.\cite{imai2007energetic,divincenzo1984p} We also considered the energy for fully intercalated lithium-graphite phase (LiC$_6$) for the AA and AB stacking structures and the AA stacking was stable while the AB was unstable as observed by earlier works.\cite{imai2007energetic,divincenzo1984p} Thus henceforth, for lithium-graphite phases, we assume that the lithium intercalated graphene layers will be AA stacked, while the non Li intercalated layers would be AB stacked.

The insertion of lithium into graphite leads to expansion of the lattice.\cite{dahn1995mechanisms} To get accurate lattice constants for each of the Li-graphite structures, we first expand graphite in-plane followed by an expansion out of plane and carry out an energy minimization.

The lithium-graphite phases have a large number of structural possibilities due to the large number of sites available for Li in graphite. A systematic way of accounting for this structural disorder is through the Ising model. 2-body cluster expansion which is similar to Ising model, has been previously used to describe the Li-graphite phases and other intercalation compounds like Li$_x$CoO$_2$, Li$_x$TiS$_2$, etc.\cite{persson2010thermodynamic,wolverton1998first,ceder2000first} In the Ising model, the Li sites are considered as a lattice model and an occupation variable is assigned to each Li site. We define the occupation variable, $s$, to be 1 if the site is occupied by Li and 0 when the site is empty. The formation enthalpy of the system is then derived in terms of these occupation variables as
\begin{equation}
\mathrm{\Delta H = C + V\sum_{i}^{}s_i + \sum_{i,j}^{}J_{i,j}s_is_j}
\label{clustexp}
\end{equation}
where s$\mathrm{_i}$ is the occupation variable associated with site `i', V is the energy associated with lithium occupying a site `i' in the graphite and J$\mathrm{_{i,j}}$ are the various interactions between two Li atoms occupying two sites in the lattice model. 

The various kinds of two-body interactions J$\mathrm{_{i,j}}$ can be associated with the different distances between the lithium occupied sites. As these Li-Li interactions are electrostatic in nature, we expect them to decrease with increasing distance. Here, we consider interactions with distance less than 5$a$ where $a$ is the distance between two adjacent carbon atoms in the graphite network. As we show later, interactions beyond this distance are negligible. For structures with stage $\mathrm{n\geq2}$ , the distance between occupied sites in different planes is greater than 5$a$ and hence there would be no out-of-plane interactions. However there will be interactions for stage 1 structures. As a result, we have two different Ising models, one for stage 1 and another for stage $\mathrm{n\geq2}$. One thing of note is that $\mathrm{\Delta H = 0}$ for graphite which implies that $\mathrm{C = 0}$ in Eq. \ref{clustexp}. Thus the final Ising models are given as:
\begin{multline}
\mathrm{\Delta H = V_i\sum_{j}^{}s_j +J_{i1}\sum_{j,k}^{}s_js_k + J_{i2}\sum_{j,k}^{}s_js_k +} \\
\mathrm{J_{i3}\sum_{j,k}^{}s_js_k + J_{i4}\sum_{j,k}^{}s_js_k  \quad for\ Stage\ n=2}
\label{inplane}
\end{multline}

\begin{multline}
\mathrm{\Delta H = V_i\sum_{j}^{}s_j + V_o\sum_{j}^{}s_j + J_{i1}\sum_{j,k}^{}s_js_k +} \\
\mathrm{J_{i2}\sum_{j,k}^{}s_js_k + J_{i3}\sum_{j,k}^{}s_js_k + J_{i4}\sum_{j,k}^{}s_js_k +} \\
\mathrm{J_{o1}\sum_{j,k}^{}s_js_k + J_{o2}\sum_{j,k}^{}s_js_k + J_{o3}\sum_{j,k}^{}s_js_k +} \\
\mathrm{J_{o4}\sum_{j,k}^{}s_js_k  \quad  for\ Stage\ n=1}
\label{outplane}
\end{multline}

\begin{figure}
\includegraphics[width=.49\linewidth]{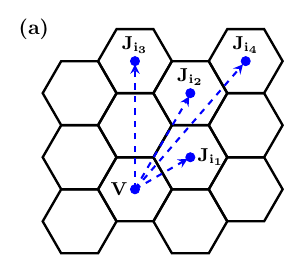}
\includegraphics[width=.49\linewidth]{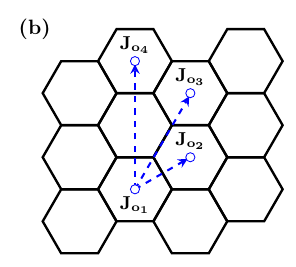}
\caption{The different Li-Li interactions in lithium intercalated graphite have been shown here. (a) shows the Li-Li interactions among the Li's in one plane i.e between the same two graphene sheets. (b) shows the interaction between a Li in a plane below the depicted plane (from where the arrows begin) with the different Li's at the depicted positions. Solid circles represent in-plane interactions and hollow circles represent out of plane interactions. These are the coefficients used in Ising model.\cite{persson2010thermodynamic}}
\label{jointer}
\end{figure}

In Eq. \ref{inplane}, J$\mathrm{_{ik}}$ represent the various Li-Li interactions within a single Li layer (in-plane) as shown in Fig. \ref{jointer}(a). In Eq. \ref{outplane}, V$\mathrm{_i}$ and V$\mathrm{_o}$  represent the Li-C interactions for the lower and upper Li layers in the stage 1 unit cells respectively, while the J$\mathrm{_{ik}}$ and J$\mathrm{_{ok}}$ represent the in-plane and out of plane interactions as shown in Fig. \ref{jointer}(b). We count all interactions in the unit cell and with neighboring unit cells with appropriate weights. For the stage 3 and stage 4 structures, out-of-plane interactions are negligible and hence neglected, which means that their energy can be calculated from the stage 2 structures as follows:
\begin{equation}
\mathrm{\Delta H_{Stage\ 3} = \frac{2}{3}\Delta H_{Stage\ 2} + H_{C_6,Stage\ 2} -  H_{C_6,Stage\ 3}} 
\label{stage3}
\end{equation}
\begin{equation}
\mathrm{\Delta H_{Stage\ 4} = \frac{2}{4}\Delta H_{Stage\ 2} + H_{C_6,Stage\ 2} -  H_{C_6,Stage\ 4}}
\label{stage4}
\end{equation}

The last two terms in Eq. \ref{stage3} and Eq. \ref{stage4} account for the energy required to reorient the graphene sheets to a different stacking. To determine the coefficients of the Ising model, we calculated 32 structures using DFT such that at least one structure had each of the interactions and all structures were unique. A regression analysis on these 32 DFT calculated structures gives all the interactions. Using the values for Ising model coefficients, the formation enthalpy for the entire phase space can be determined. We explore the phase space through various unit cell sizes of the graphite where we intercalate lithium atoms and calculate energies using the Ising model. We could explore unit cells with a maximum of 24 Li sites for stage 1 and stage 2, beyond which the combinations of filling Li atoms become very large ($\mathrm{\sim2^{24}}$). Thus we can only explore upto the phase Li$\mathrm{_{0.03}}$C$_6$.  Ordered Li$_x$C$_6$ phases with $\mathrm{x < 0.03}$ have not been observed experimentally due to the formation of SEI during the initial electrochemical lithium intercalation.\cite{reynier2003entropy} 

Christensen {\it et al.} have shown that systematic errors in the formation energy of alkali oxides can be reduced through the use of a reference compound that has the same oxidation state.\cite{christensen2015identifying} Thus for this work, we correct for the value of the reference energy of lithium by the adding the term shown in the following equation,
\begin{equation}
\mathrm{\Delta H_{Li} = H_{LiC_6} - H_{C_6} - \Delta H_{LiC_6}^{f, exp}}
\end{equation}
where $\mathrm{H_{LiC_6}}$, $\mathrm{H_{C_6}}$, $\mathrm{H_{Li}}$ are DFT calculated energies of LiC$_6$,  C$_6$, and Li and $\mathrm{\Delta H_{LiC_6}^{f, exp}}$ is the experimental  formation enthalpy of LiC$_6$.

\subsection{Phonon Calculations}
The phonon calculations were implemented using the density functional perturbation theory (DFPT)\cite{baroni2001phonons} in Quantum Espresso.\cite{giannozzi2009quantum} The calculations were done using ultra-soft pseudopotentials. The lattice parameters were taken from the GPAW calculations and re-optimized in Quantum Espresso\cite{sabatini2012structural} before doing the phonon calculations. The optB88-vdW\cite{klimevs2009chemical} exchange correlation functional which employs the vdW-DF\cite{dion2004van,thonhauser2007van,thonhauser2015spin,berland2015van} non local correction was used for phonon calculations because the phonon dispersion curves agree very well with experiments.\cite{hazrati2014li} For the phonon calculations, the structures were further relaxed until the forces on atoms are less than 1 meV/\AA. A kinetic energy cutoff of 550 eV is employed for the plane wave expansion of the wave-functions and a 5500 eV cutoff was chosen for charge density and potential. The same k-point grid mentioned earlier was used for the phonon calculations. The phonon density of states were calculated for Li, C$_6$, LiC$_6$, LiC$_8$, LiC$_{12}$ and LiC$_{18}$. These were the used to calculate the finite temperature vibrational contributions to enthalpy and entropy under the harmonic approximation as given below.
\begin{equation}
\mathrm{H_{vib}(T) =  \int_{0}^{\infty} g(\omega)d\omega \Bigg(\frac{1}{2}h\omega + \frac{h\omega}{e^{h\omega/k_BT}-1}\Bigg)}    
\end{equation}
\begin{equation}
\mathrm{S_{vib} = k_B \int_{0}^{\infty} g(\omega)d\omega \Bigg(\frac{h\omega/k_BT}{e^{h\omega/k_BT}-1} + ln\Big(1-e^{h\omega/k_BT}\Big)\Bigg)}    
\end{equation}
In the above expressions, g($\omega$) are the normalized phonon density of states. The zero point energy is included in H$\mathrm{_{vib}}$(T) as the $\frac{1}{2}$h$\omega$ term. To get the vibration contributions for all lithium-graphite phases, a fourth order polynomial function was fit to the vibrational contribution to formation enthalpy and entropy as a function of the lithium fraction in the phases. 

\subsection{Entropy Calculation}

The entropy change associated with the formation of the Li-graphite phase can be written as $\mathrm{\Delta S = S_{Li_xC_6} - S_{C_6}- xS_{Li}}$. Entropy of a material consists of the configurational entropy and the entropy of the vibrational modes of the structure. The entropy change is now given by: 
\begin{equation}
\mathrm{\Delta S = S_{Li_xC_6}^{conf.} + S_{Li_xC_6}^{vib.} - S_{C_6}^{vib.} - xS_{Li}^{vib.}}
\end{equation}

\subsubsection{Configurational entropy}
To derive the configurational entropy, we can consider a graphite block with N atoms. A graphite block of N atoms due to its hexagonal symmetry will have N/2 sites for lithium to occupy. From the interaction coefficients, we can see that the first in-plane interaction coefficient is significantly larger than the others, which implies that it is very unfavorable for two lithium atoms to occupy adjacent sites. Hence due to the hexagonal symmetry of graphite, we can only fill a Li among 3 adjacent sites. This implies that for a particular phase x of Li$_x$C$_6$, the total configurations $\mathrm{\Omega}$ of possible Li ordering is the number of ways of arranging $\mathrm{\frac{Nx}{6}}$ lithium atoms in $\mathrm{\frac{N}{6}}$ sites. This is valid for a stage 1 compound where the Li can occupy every layer of graphite. Now to include staging, we have to reduce the number of lithium sites as every layer will not be occupied. Based on the definition of staging, a stage $n$ compound will have $\mathrm{\frac{N}{6n}}$ sites. Thus $\mathrm{\Omega = \binom{\frac{N}{6}}{\frac{Nx}{6n}}}$. The configurational entropy for Li$_x$C$_6$ for N carbon atoms can now be written as follows:
\begin{equation}
\mathrm{S_{conf} = k_B ln(\Omega)}
\end{equation}
For our model, we need to normalize the configurational entropy to 6 carbon atoms. i. e. $\mathrm{s_{conf} = \frac{6}{N} S_{conf}}$. 
Now we can derive the expression for configurational entropy of phase x as given by:
\begin{equation}
\mathrm{s_{conf} = \frac{6k_B}{N} ln\Bigg( \binom{\frac{N}{6}}{\frac{Nx}{6n}}\Bigg)}
\end{equation}
After applying the Stirling's approximation for large N, the configurational entropy for a stage $n$ compound can be expressed as:
\begin{equation}
\mathrm{s_{conf} = -k_B \Bigg( xln(x) + \frac{1}{n}ln\Big(\frac{1}{n}\Big) +  \Big(\frac{1}{n}-x\Big)ln\Big(\frac{1}{n}-x\Big) \Bigg)}
\end{equation}

\subsubsection{Vibrational entropy}
 The vibrational entropies for the different Li-graphite phases was calculated using the phonon density of states as described in the previous section. 

\subsection{Uncertainty Estimation}
The uncertainty estimates follow the standard BEEF ensemble estimation procedure.\cite{medford2014assessing,deshpande2016quantifying} BEEF generates an ensemble of energies for all the 32 DFT calculated structures. These ensemble of energies represents the energies calculated from an ensemble of 2000 exchange correlation functionals and hence determines the DFT uncertainty for energy. This uncertainty is propagated in the Ising model by determining the Ising model coefficients for every functional through regression, resulting in an ensemble for each coefficient. For every functional in the ensemble, we then calculate the energies for all possible structures using the Ising model. After adding the contributions from phonons and configurational entropy, we determine the convex hull of phases and in turn the voltage diagram for each functional in the ensemble. Each ensemble will give a unique phase diagram. For each of the identified phases in all functionals, we can determine the confidence of determining the phase as a stable phase. A stable phase is the phase with minimum Gibbs free energy for a given $x$. Recently, the concept of confidence value was defined in context of magnetic ordering of materials.\cite{houchins2017quantifying} Similarly, we define the confidence value of a phase can be defined as the fraction of BEEF ensemble of functionals that predict the phase as a stable phase for a given $x$. The confidence value (c) of phase $s_i(x)$ is given by the following equation
\begin{equation}
\mathrm{ c(s_i(x)) = \frac{1}{N_{ens}}\sum_{n=1}^{N_{ens}} \delta (\Delta G_n (s_i(x)) - min_{s_j \epsilon S}\{\Delta G_n (s_j(x))\}) }      
\end{equation}

where $N_{ens}$ is the number of functionals in the BEEF ensemble, S is the set of all phases $s_j$ and mixtures of other phases at a given $x$, $\Delta G_n(s_i(x))$ is the Gibbs free energy of formation of phase $s_i(x)$ calculated using the $n^{th}$ functional in the BEEF ensemble and $\delta$ is the Dirac delta function.  If all functionals predict a phase to be stable, the corresponding confidence value is 1, while no functional predicts a phase to be stable, the confidence value is 0.  When a large fraction of functionals predict the same stable phase for a given $x$, then there is a high confidence that the prediction is independent of exchange correlation functional and is likely to be the qualitatively correct.  Such an approach has been utilized successfully to identify majority of the stable phases of Li-Ni-Co-Mn oxide.\cite{houchins2018towards}

Neglecting the uncertainty in phonon calculations, this method gives us the exact uncertainty estimates for the lithium-graphite phase diagram predicted from GGA level DFT calculations. 

\section{Results}

The formation enthalpies for 32 structures have been calculated using DFT as per the procedure described in the Methods section. Table I in the S.I. provides the different kind of interactions, the exact Li$_x$C$_6$ phase, the lattice constants for each of the structures, the stage number for the 32 structures considered. The ZPE is added later in the phonon contribution to enthalpy.

\begin{figure}
\includegraphics[scale=0.3]{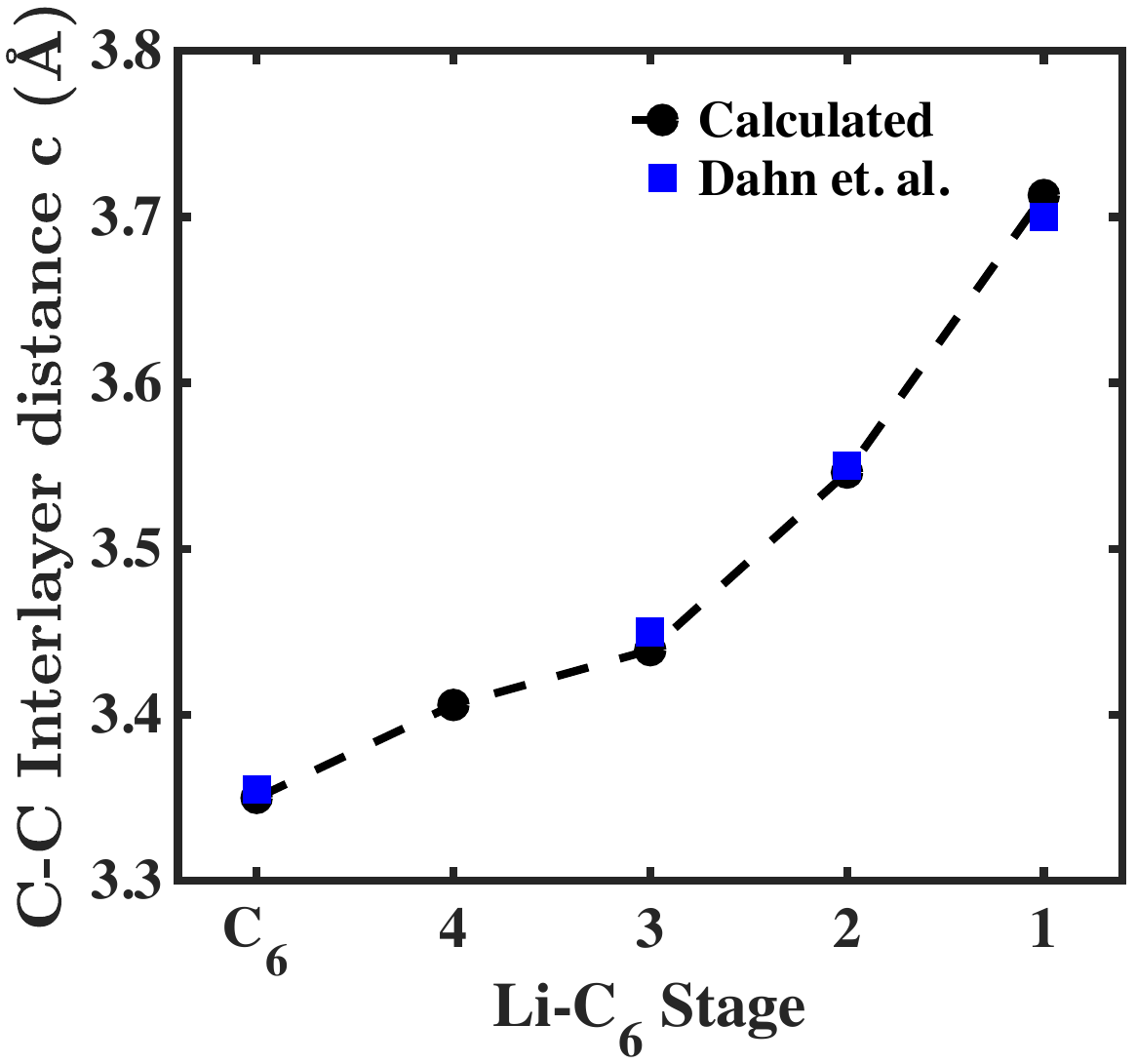}
\caption{The increase in the C-C inter-layer distance among adjacent graphene sheets as a functions of the various stages of lithium intercalation. The black dots are the DFT calculated values and blue dots are experimental values by Dahn et. al.\cite{dahn1990suppression} Note that the inter-layer distance increases non linearly proving that the vdW interactions get screened non uniformly upon intercalation.} 
\label{cvs}
\end{figure}

The in-plane lattice constant $a$ increases slightly upon lithium intercalation. However, out-of-plane lattice constant, $c$, which is the distance between adjacent graphene sheets significantly increases upon lithium intercalation to a maximum increase of 10.8$\%$ upon full intercalation. $c$ is more strongly dependent on the stage number of Li$_x$C$_6$, rather than the filling fraction, $x$. A plot of $c$ as a function of the stage is shown in Fig. \ref{cvs} and agrees very well with experiments by Dahn et al.\cite{dahn1990suppression} This implies that the upon intercalation, the Li is weakening the vdW interactions between the graphene planes.  This plot shows that the weakening effect is non-linear and hence, a constant vdW correction as assumed by Persson {\it et al.}\cite{persson2010thermodynamic} may be prone to errors.

Using the kind of interactions involved in each structure along with its formation energy, we determined the interaction coefficients for the Ising model for stage 1 and stage n $\geq$ 2. We use the least squares regression method to evaluate the interaction coefficients. The maximum error for the fit for stage 1 Ising model is 0.015 eV and for stage n $\geq$ 2 Ising model is 0.013 eV. These errors are maximum deviations of the Ising model from the DFT energies after performing the least squares method analysis to determine Ising model coefficients using MATLAB software. The evaluated interaction coefficients are shown in Table I.
\begin{table}
\caption{\label{intertable1}The evaluated Ising model coefficients for stage 1 and stage 2 compounds along with the corresponding errors. All the values are in units of eV and have been normalized for 6 C atoms. Note that the stage 2 Ising model does not have out-of-plane interactions, hence no values for J$\mathrm{_{o_k}}$ have been reported.}
\begin{tabular}{|c|c|c|c|c|c|c|c|c|c|c|c|}
\hline
& $V_i$ & $V_o$ & $J_{i1}$ &  $J_{i2}$ & $J_{i3}$ & $J_{i4}$ & $J_{o1}$ & $J_{o2}$ & $J_{o3}$ & $J_{o4}$ \\
\hline
n = 2& $-0.49$ & $-$ & $0.41$ & $0.11$ & $0.1$ & $0.01$ & $-$ & $-$ & $-$ & $-$ \\
\hline
$1\sigma$& \multicolumn{1}{|r|}{$0.34$} & $-$ & $0.15$ & $0.08$ & $0.06$ & $0.01$ & $-$ & $-$ & $-$ & $-$ \\
\hline
n = 1& $-0.36$ & $-0.21$ & $0.32$ & $0.05$ & $0.05$ & $0.00$ & $0.01$ & $0.03$ & $0.02$ & $0.00$ \\
\hline
$1\sigma$ & \multicolumn{1}{|r|}{$0.28$} & \multicolumn{1}{|r|}{$0.17$} & $0.18$ & $0.05$ & $0.06$ & $0.01$ & $0.02$ & $0.06$ & $0.02$ & $0.00$ \\
\hline
\end{tabular}
\end{table}

\begin{figure}
\includegraphics[scale=0.3]{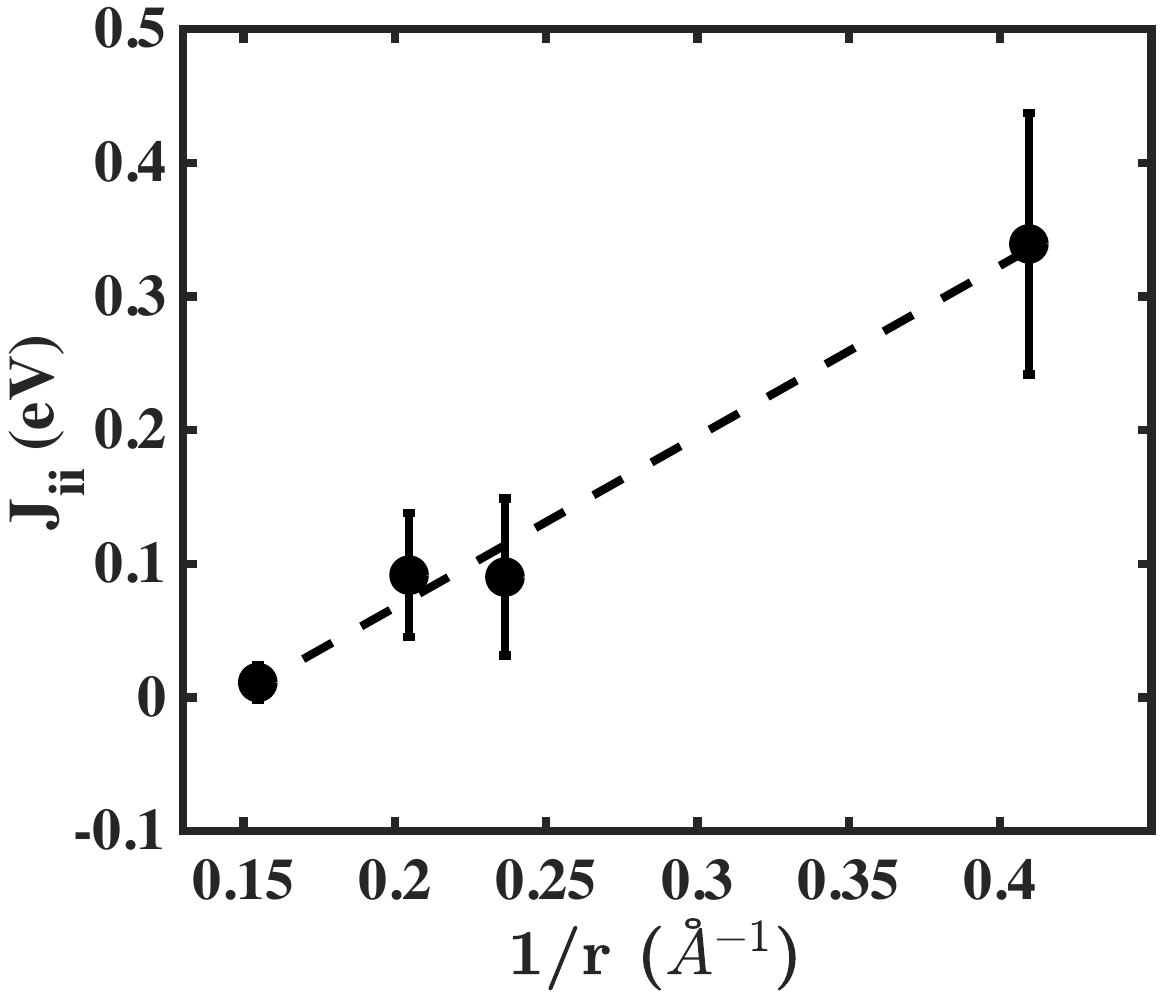}
\caption{The in-plane Li-Li interactions plotted as a function of the inverse of the distance between the two Li's involved in the interaction. The interactions vary linearly with 1/r and the best fit line shown has an $R^2 = 0.99$.}
\label{jvr}
\end{figure}

As we can observe from the table, the occupation terms, $\mathrm{V}$, which correspond to Li-C interactions are negative implying that these interactions are attractive while the $\mathrm{J_{ii} 's}$ and the $\mathrm{J_{oi} 's}$ are positive implying that the Li-Li interactions are repulsive in nature.  The in-plane Li-Li interactions ($\mathrm{J_{ii}}$) are plotted against the inverse of the corresponding distances of the interactions and shown in Fig. \ref{jvr}. The in-plane Li-Li interactions decrease with increasing distance and are proportional to 1/r showing that the interactions are dominantly electrostatic in nature. As can be seen from Table, the out-of-plane Li-Li interactions ($\mathrm{J_{oi}}$) are small and do not follow a similar trend with distance associated with interactions. We attribute this to varying degrees of charge screening by the graphene sheets for these interactions.

We use the procedure described in the methods section to calculate the formation enthaply for different unit cell sizes up to 24 Li sites for different phases of Li$_x$C$_6$ and different stages. We calculated formation enthalpies for about 1 billion different structures. To determine the Gibbs free energy of structures, we also need to add  configurational entropy, vibrational enthalpic and entropic contributions at finite temperature as described earlier.

\begin{figure}
\includegraphics[scale=0.42]{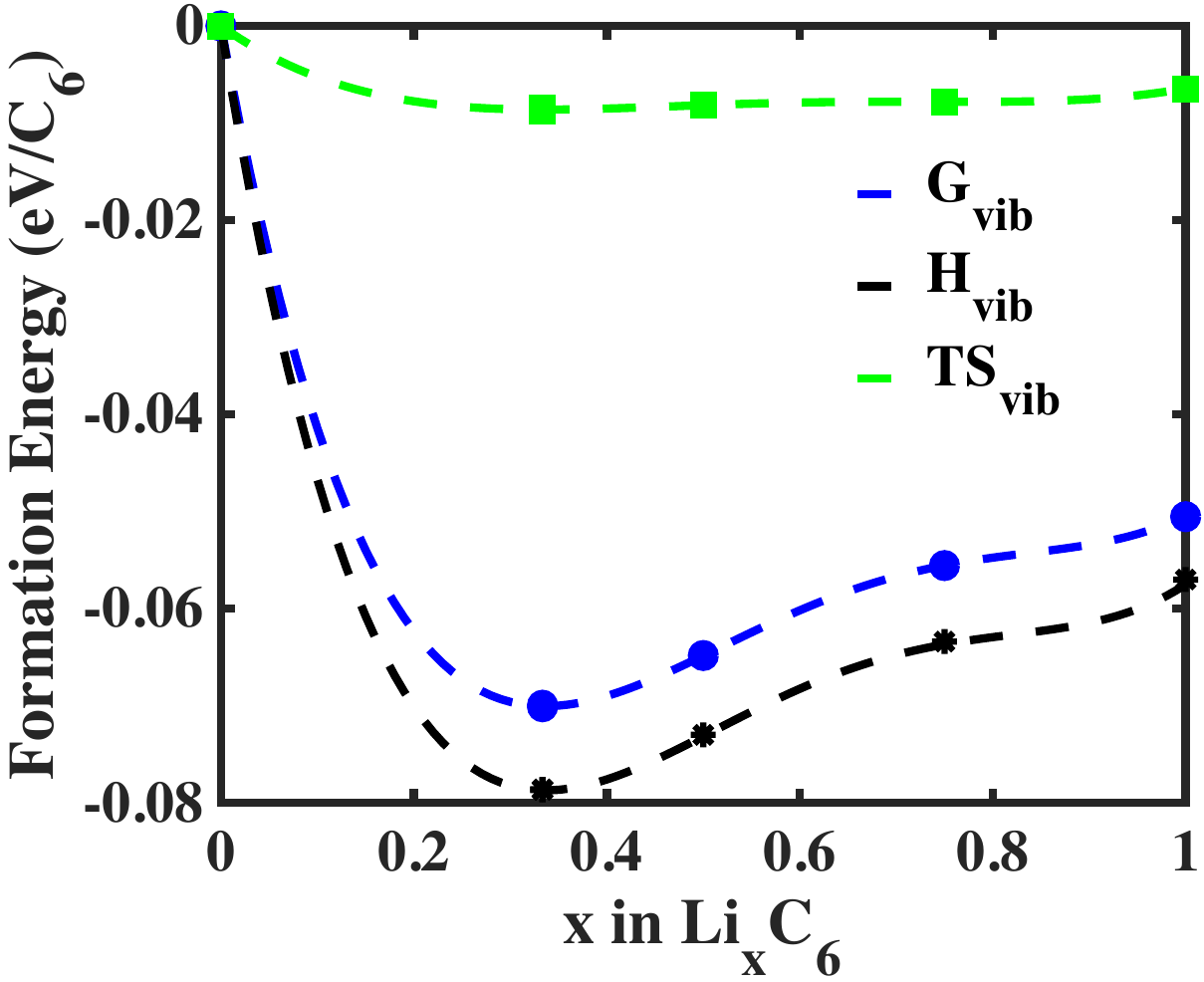}
\caption{Phonon contributions to formation enthalpy, entropy and Gibbs free energy at T = 298 K are shown in black, green and blue colors respectively. The points are the calculated values from DFT and the dotted lines are the fitted 4$^{th}$ order polynomial functions. }
\label{svx}
\end{figure}

 The configurational entropy is dependent on stage and lithium fraction as described in the methods and has significant contribution of $\sim$0.02 eV at T = 298 K for phases x$<$0.25 in Li$_x$C$_6$, but it is $<$0.01 eV for phases x$>$0.25. The vibrational contribution to free energy due to phonons is shown in Fig. \ref{svx}. We see that the vibration entropy contribution is small and negative, about $\sim$ -5 meV. We can attribute this loss of entropy to the hindered motion of the lithium atoms in the intercalated phases as compared to bulk lithium. The total entropy of a phase is the sum of configurational and vibrational entropy.  Thus, we can see that the the entropy is positive for phases x$<$0.25 but negative for x$>$0.25 which agrees qualitatively with measurements done by Reynier {\it et al.}\cite{reynier2003entropy} In the work by Reynier {\it et al.}, the entropy change was calculated by measuring the change in open circuit voltages in half cells with graphite electrodes for a small change in temperature. We also see that the enthalpic contribution due to phonons is significantly large about -0.08 eV and definitely affects the phase diagram. We find that the $\Delta$ZPE dominates the enthalpic contribution due to phonons. The $\Delta$ZPE contribution is negative as the high frequency modes of graphite shift to lower frequencies upon intercalation as seen Fig 1. in S.I. 

\begin{figure}
\includegraphics[width=1.0\linewidth]{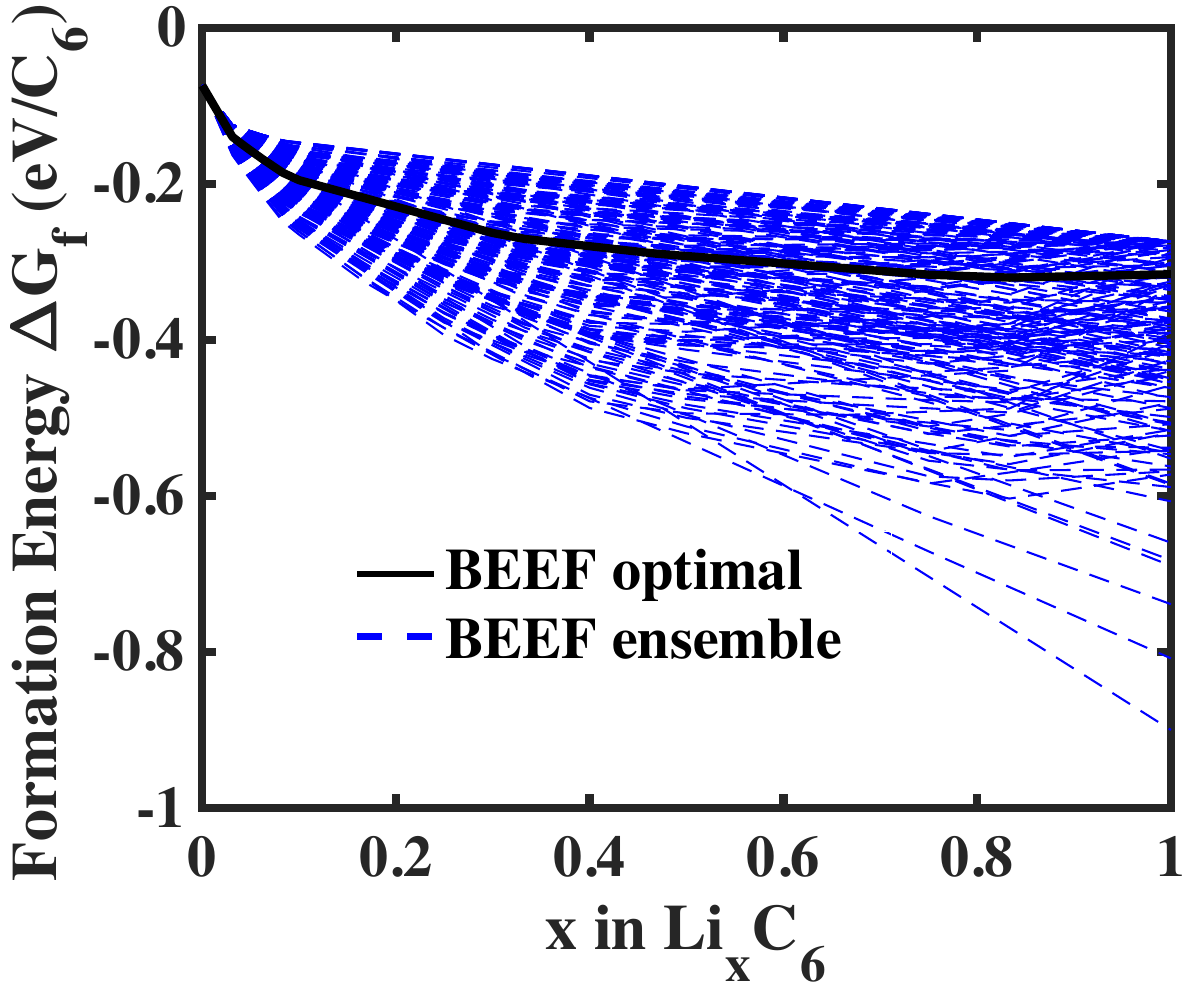}
\caption{The ensemble of convex hulls of $\Delta$G vs. x which represents the free energy phase transformation diagram for Li intercalation in graphite for the 2000 BEEF ensemble of functionals. The blue dotted lines show the ensemble while the black solid line corresponds to the BEEF optimal functional.}
  \label{gcx}
\end{figure}

To construct the phase diagram of Li-graphite, we need to choose the structures which lie on the convex hull of free energy vs. $x$ in Li$_x$C$_6$ diagram. The phase diagram being the convex hull is a result of the condition that a system in thermodynamic equilibrium minimizes its free energy. The convex hull is derived from the 1 billion points which gives the phase diagram as shown in Fig. \ref{gcx}. From the convex hull, we evaluate the intercalation potentials, using the free energy of the most stable phases. The intercalation potential diagram is shown in Fig. \ref{Vvx}. The calculated intercalation potentials are within 0.05 V of the experimentally measured potentials for x$>$0.3 but differ by 0.1-0.4 V for x$<$0.3 also shown in Fig. \ref{Vvx}. The experimental results are from very slow lithiation and delithiation of graphite at C/100 current rates. The disagreement in potential for low $x$ phases might be from a number of factors including errors from phonon contributions, SEI formation in experiments, errors from Ising model, etc. The energy differences for these low $x$ phases are quite small, which means that the potentials are more sensitive to errors. However, it is worth pointing out that we are able to predict all of the phase transformations observed in experiments accurately. The BEEF ensemble of energies was used to propagate uncertainty in the Ising model and phase diagrams. The ensemble of convex hulls for Gibbs free energy vs $x$ is shown in Fig. \ref{gcx}. We see that a large number of GGA functionals predict similar convex hulls as the BEEF optimal functional. However, there are several functionals which predict vastly different stable phases and widely varying intercalation potentials. This shows that the choice of functional is very important in predicting phase diagrams and raises an important question of how robust are the identified phases and intercalation potentials, which is discussed below.

\begin{figure*}
\includegraphics[scale=0.4]{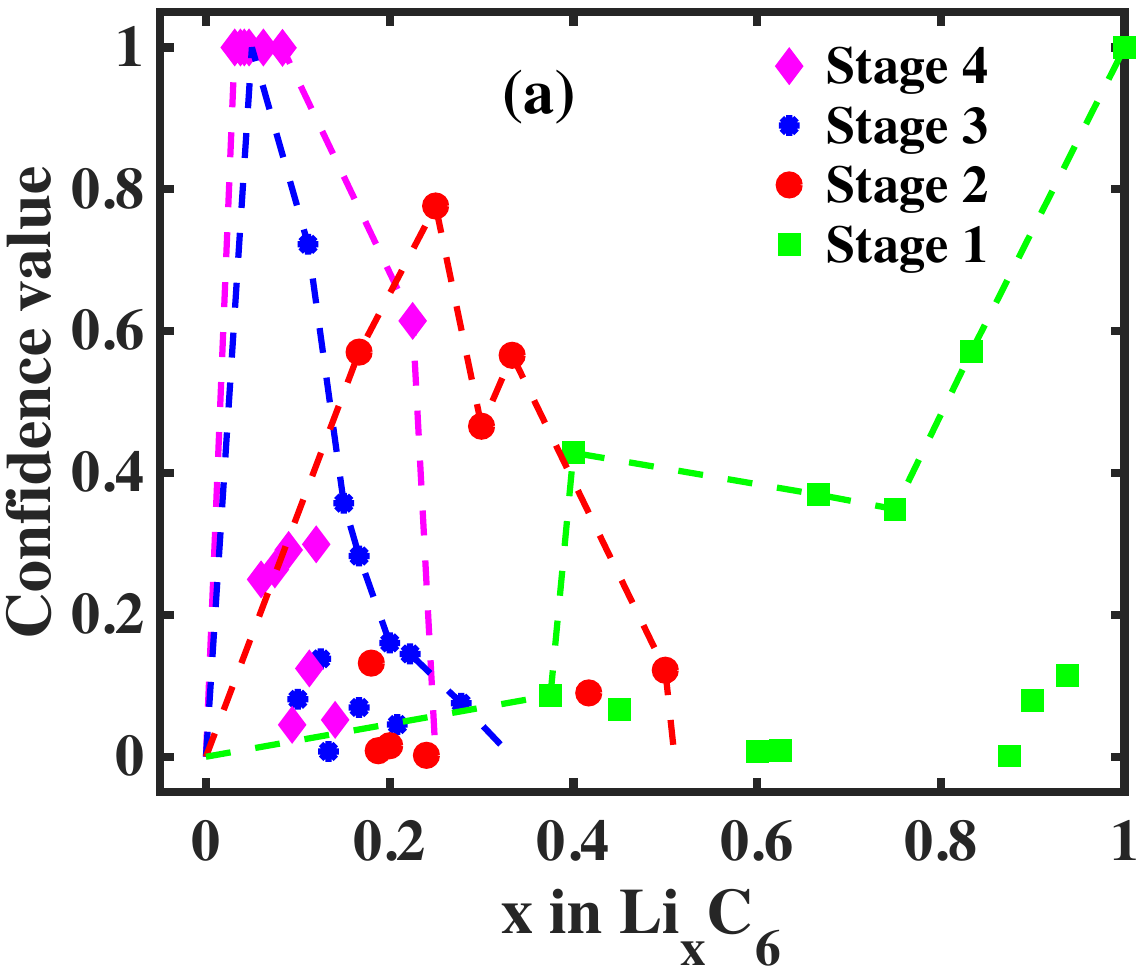}
\hspace{1cm}
\includegraphics[scale=0.4]{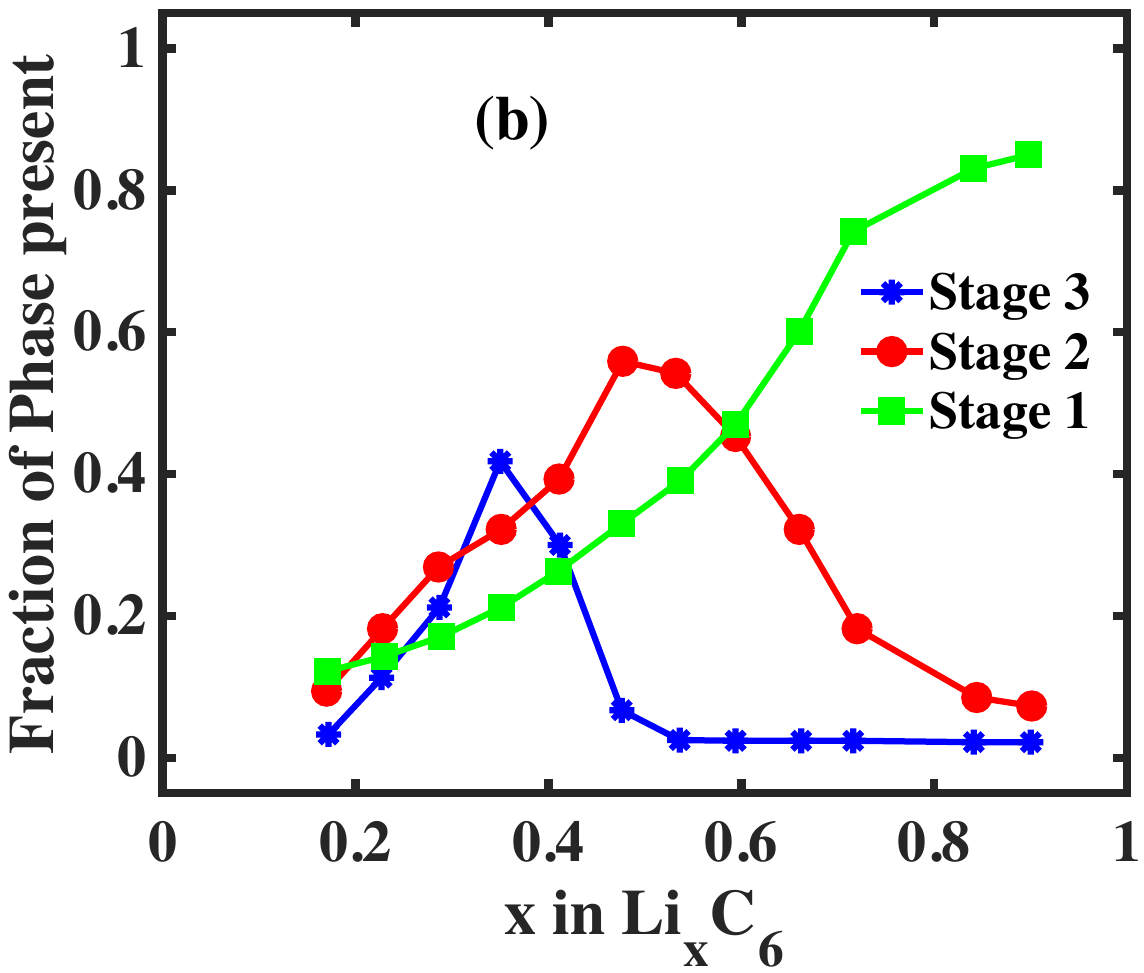}
\caption{(a) The confidence value of phases with different staging as function of lithium content $x$. The dots are all the stable phases determined by the ensemble. The lines have been drawn to only connect the high confidence value phases(b) The experimentally measured fraction of phases with different staging at a given $x$ by Dahn et. al.\cite{dahn1990suppression}}

\label{confidence_expt}
\end{figure*}

\begin{figure}
\includegraphics[scale=0.36]{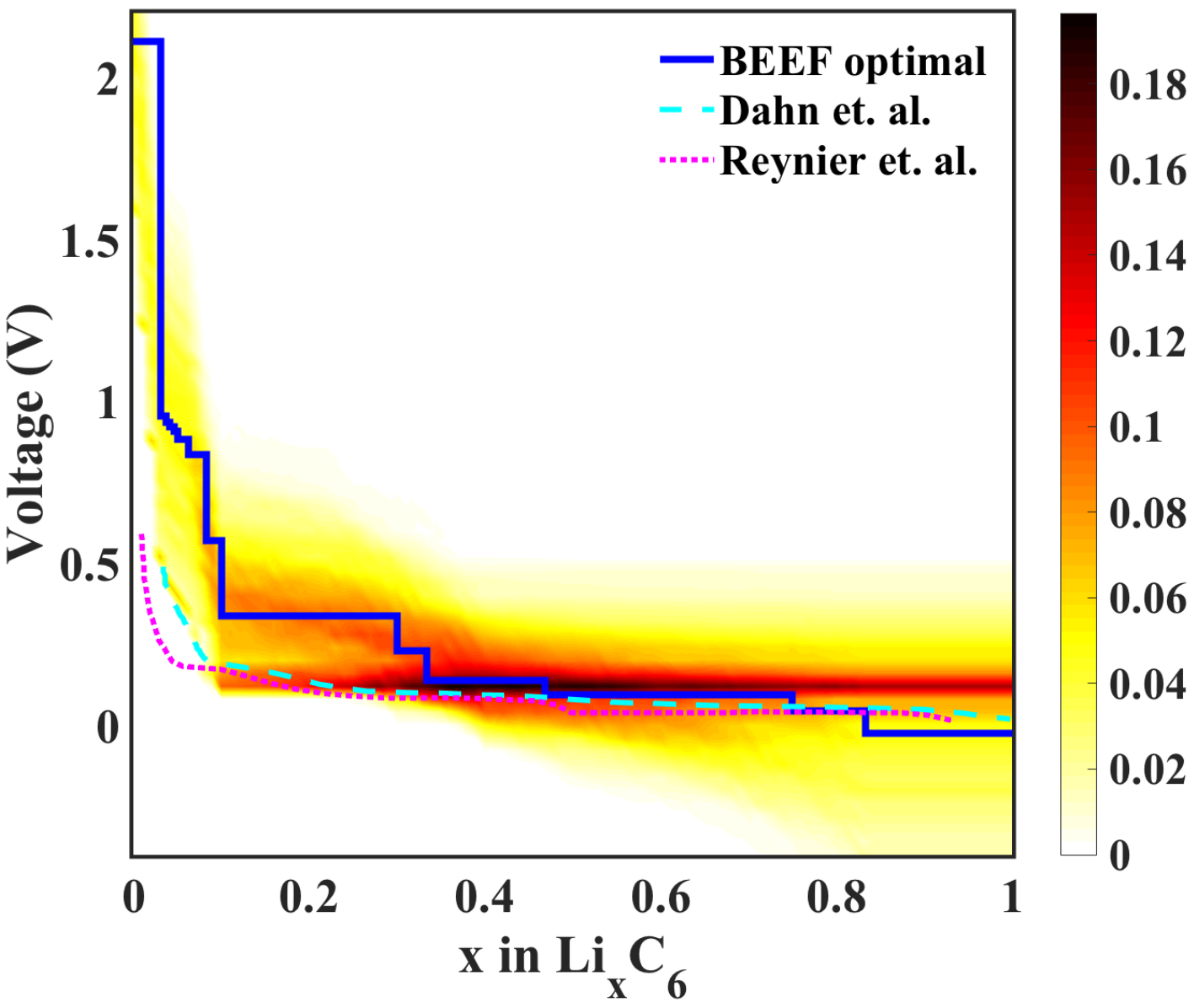}
\caption{The interacalation/phase transformation potential for stable phases as function of x in Li$_x$C$_6$. The blue line represents the BEEF  intercalation potential calculated using the BEEF optimal functional. The contour map represents the probability density function of intercalation potential at each $x$ for the BEEF ensemble. The larger the probability, the more the number of functionals that predict that potential value. The cyan and pink lines represent the experimentally observed intercalation potential diagrams by J. Dahn {\it et al.} and Y. Reynier {\it et al.}\cite{dahn1991phase,reynier2003entropy}}
\label{Vvx}
\end{figure}

The BEEF optimal functional predicts a phase diagram where there are about 10 stable stage 4 phases for x$<$0.1. This is followed by a phase transformation to  stage 3 $x=0.1$ phase. The next stage transformation is to stage 2 compounds $x=0.3$, $x=1/3$, $x=0.4685$ and $x=0.5$. This is followed by transformation to stage 1 compounds $x=0.75$, $x=5/6$ and $x=1$. To determine the uncertainty on these phases, we determine fraction functionals from the BEEF ensemble actually predict these as the stable phases, which is the confidence value of these phases as defined in the methods section and shown in Table II. We see that the confidence value for $x<0.1$ is 1, which means we predict this region to solely to comprising of stage 4 compounds. For $0.1<x<0.3$, the confidence value is very low, which suggests that this region is most likely comprising of a mixture of stage 4, stage 3 and stage 2 compounds. For $0.3<x<0.5$, the confidence value is slightly more than 0.5 which indicates that the compounds are likely stage 2 and stage 1 compounds, but there may be some stable phases in addition to ones predicted by BEEF optimal. For $0.5<x<1$, the confidence value is low indicating that this region is probably is mixture of stage 1 compounds. Our ensemble of functionals do capture all stable phases from previous studies such as $x = 0.1667, 0.25, 0.3, 0.3333, 0.375, 0.5, 0.875$ and 1.\cite{persson2010thermodynamic,hazrati2014li,filhol2008phase,okamoto2013density} We also assign a confidence value for each phase and observe that $x = 0.375,0.5,0.875$ have a low confidence value implying that there might be mixtures of other stable phases more stable at that composition.  All these phases and mixture of phases at a given $x$ agree very well with those observed in experiments by X-ray diffraction,\cite{dahn1991phase} and optical measurements\cite{guo2016li}. In Fig. \ref{confidence_expt}, we also compare the confidence value with the experimentally measured fraction of phases present at a given $x$ and see that there is a qualitative similarity in the data. We observe similar coexistence of compounds of different stages and that the experimental data might be mixtures of the phases with confidence value $>$0.25. The threshold confidence value does not have a firm basis and needs further basis. The experimental data might have some error in $x$ data due to formation of SEI.\cite{dahn1990suppression} Thus, a low confidence value can be used as an indicator of existence of multiple stable structures at given $x$, i.e. a solid solution.

The ensemble of intercalation potentials at each $x$ was used to create a probability density function (pdf). The ensemble gives 2000 predicted potentials at each $x$. A histogram for these 2000 values with 0.01 V bin width is then used to create a pdf at that $x$. This pdf is plotted as the contour in Fig. \ref{Vvx}. The pdf gives us the robustness in intercalation potential of GGA level DFT. As shown in Fig. \ref{Vvx}, the probability of predicting the correct voltage increases with $x$. Also the pdf maximum near $x=1$ is slightly different than the BEEF optimal but is actually very close to the experiment. This suggests that likelihood of predicting the correct intercalation potentials could be much higher using multiple functionals than a single functional. A similar idea of using an adaptively weighted sum of energies from multiple functionals to get better agreement with experiments has been shown for adsorption energies.\cite{hensley2017dft} This can be attributed to the fact that use of multiple functionals can accurately determine a larger variety of energetic interactions and bonding environments.

\begin{table}
\caption{The confidence values for all phases predicted by the BEEF optimal functional. Confidence value is the fraction of BEEF ensemble of functionals predicting a phase as a stable phase. $*$ indicates the phases with confidence value $>$0.25, which were not predicted by the BEEF optimal functional.}

\begin{tabular}{|c|c|c|}
\hline
x	&	Stage	&	Confidence value	\\
\hline
0	&	4	&	1	\\
\hline
0.0313	&	4	&	1	\\
\hline
0.0375	&	4	&	1	\\
\hline
0.0417	&	4	&	1	\\
\hline
0.0469	&	4	&	1	\\
\hline
0.05	&	3	&	1	\\
\hline
0.06*	&	4	&	0.2500	\\
\hline
0.0625	&	4	&	1	\\
\hline
0.075*	&	4	&	0.2640	\\
\hline
0.0833	&	4	&	0.9995	\\
\hline
0.09*	&	4	&	0.2915	\\
\hline
0.1	&	3	&	0.081	\\
\hline
0.1111*	&	3	&	0.7225	\\
\hline
0.12*	&	4	&	0.2995	\\
\hline
0.15*	&	3	&	0.3575	\\
\hline
0.1667*	&	3	&	0.2830	\\
\hline
0.1667*	&	2	&	0.5705	\\
\hline
0.225*	&	4	&	0.6145	\\
\hline
0.25*	&	2	&	0.7765	\\
\hline
0.3	&	2	&	0.466	\\
\hline
0.3333	&	2	&	0.566	\\
\hline
0.4*	&	1	&	0.4285	\\
\hline
0.4688	&	2	&	0.122	\\
\hline
0.5	&	2	&	0.085	\\
\hline
0.6667*	&	1	&	0.3695	\\
\hline
0.75	&	1	&	0.349	\\
\hline
0.8333	&	1	&	0.5715	\\
\hline
1	&	1	&	1	\\
\hline
\end{tabular}
\end{table}

We also calculated the temperature dependence of the phase diagram for the BEEF optimal functional from -25$^{\circ}$C to 50$^{\circ}$C. We find that in this temperature range, we do not observe any change in the phase diagram and the final intercalation potential plateau decreases by 0.01 V as shown in Fig 2 in S.I. This small change in voltage corresponds to only a 1\% change in the onset of lithium plating, corresponding to voltage of 0. However, higher amount of lithium plating has been observed experimentally at cold temperatures ($\sim$-20$^{\circ}$C).\cite{bugga2010lithium} This suggests that lithium plating is probably dependent on kinetics and not thermodynamics. 

The identification of different stable structures at a given $x$ enables calculations for getting a better understanding of lithium diffusion in graphite. The stage 4 and stage 3 compounds that exist at low $x$ phases determine the initial kinetic barrier for lithium intercalation in graphite and hence should be studied more carefully. The transport in these compounds may help us to understand and decouple other kinetic effects from SEI and anode microstructure.  Finally, we believe the approach of utilizing uncertainty quantification to robustly determine the stable phases will be broadly important for determining phase diagrams of Li-ion anodes and cathodes.

\section{Summary and Conclusion}
The presented framework comprising of DFT and statistical thermodynamics accurately predicts the phase diagram and intercalation potentials for electrochemical Li intercalation in natural graphite. We highlight the importance of accurately capturing vdW interactions and zero point energies to determine the stable phases during intercalation. We introduced a method to evaluate uncertainty of free energies of various phases and the associated intercalation potentials through the built-in error estimation capability of the BEEF-vdW exchange correlation functional. We have derived confidence values for different phases, which represents the uncertainty associated with DFT predicting them as stable phases.   We also associate the confidence value with the fraction of phases existing at given $x$ and get a good agreement with experiments. The confidence value we believe would be a good indicator to identify the disorder at a given concentration in phase diagrams. We also find that predicting intercalation potentials from a statistical approach using multiple functionals will give us more robust predictions and may yield better agreement with experiments. Lastly we determine that the lithium-graphite phase diagram is not sensitive to changes in temperature, suggesting that the dominant effect triggering lithium plating in lithium ion cells is not thermodynamic. This high fidelity, large scale prediction of lithium-graphite phase diagram incorporating uncertainty and other important corrections will be important for next-generation robust battery material design.

\begin{acknowledgments}
 The authors wish to acknowledge helpful discussions with Jay Whitacre and Shawn Litster.  Acknowledgment is made to the Scott Institute for Energy Innovation at Carnegie Mellon for partial support of this research. This work was also supported in part by the Pennsylvania Infrastructure Technology Alliance, a partnership of Carnegie Mellon, Lehigh University and the Commonwealth of Pennsylvania’s Department of Community and Economic Development (DCED).  
 \end{acknowledgments}

\bibliography{Graphite_ref.bib}

\newpage
\includepdf[pages={{},{},1,{},2,{},3,{},4,{},5,{},}]{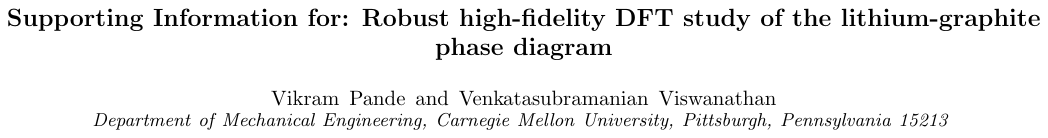}\AtBeginShipout\AtBeginShipoutDiscard

\end{document}